 \documentclass[final,5p,times,twocolumn]{elsarticle} % authoryear

\usepackage{amssymb}
\usepackage{lipsum}
\usepackage{amsmath}
\usepackage{bm}
\usepackage{hyperref}
\usepackage{xcolor}
\biboptions{sort&compress}

\begin{document}

\begin{frontmatter}

\title{Principal Components of Nuclear Mass Model Residuals}

\author[FZU]{Y. Y. Huang}

\author[FZU]{X. H. Wu\corref{cor1}}
\ead{wuxinhui@fzu.edu.cn}

%\date{\today}

\cortext[cor1]{Corresponding Author}
\address[FZU]{Department of Physics, Fuzhou University, Fuzhou 350108, Fujian, China}
%\address[PKU]{State Key Laboratory of Nuclear Physics and Technology, School of Physics, Peking University, Beijing 100871, China}

\begin{abstract}
Principal Component Analysis (PCA) is applied to the residuals of six widely used nuclear mass models to uncover systematic deviations and identify missing physical effects in theoretical nuclear mass predictions. 
By analyzing the principal components of nuclear mass model residuals, this study reveals that no single dominant pattern governs the discrepancies across models. 
Instead, the residual structures are largely uncorrelated, indicating that current nuclear mass models fail to capture underlying nuclear residual effects in distinct and model-specific ways.
These findings suggest that improvements to nuclear mass models should be guided by model-specific residual analyses rather than a one-size-fits-all approach.
Model-specific residual patterns are suggested for six nuclear mass models respectively.
\end{abstract}


\end{frontmatter}

%\tableofcontents

%-----------------------------------------------------
\section{Introduction}
%-----------------------------------------------------
Nuclear masses are fundamentally important for nuclear physics, as they can reflect many underlying physical effects related to nuclear structure~\cite{Lunney2003Rev.Mod.Phys., Zhang2025AAPPSBulletin}.
Nuclear masses are also significantly important for astrophysics, as they determine the reaction energies that go into the calculations of all involved nuclear reaction rates in the stellar evolutions~\cite{Mumpower2016Prog.Part.Nucl.Phys., Jiang2021Astrophys.J., Wu2022Astrophys.J.152, Huang2025ApJ}.
Great achievements in nuclear mass measurements have been recently made thanks to the development of radioactive ion beam facilities, and about 2500 nuclear masses have been measured to date~\cite{Wang2021Chin.Phys.C}.
Nevertheless, there is still a large uncharted territory in the nuclear landscape that cannot be accessed experimentally even in the foreseeable future.

Theoretical prediction of nuclear properties is an extremely tough challenge, due to the difficulties in tackling both nuclear interactions and quantum many-body systems.
To accurately describe nuclear masses, one should in principle properly address all the underlying effects of nuclear quantum many-body systems, e.g., bulk effects, deformation effects, shell effects, odd-even effects, and even some unperceived effects.
Nuclear mass predictions can be traced back to the macroscopic Weizs\"{a}cker mass formula based on the liquid drop model (LDM)~\cite{Weizsaecker1935Z.Physik}, which includes the bulk properties of nuclei quite well but lacks other effects.
Efforts have been made to include more and more effects by developing  macroscopic-microscopic models~\cite{Wang2014Phys.Lett.B, Moeller2016Atom.DataNucl.DataTables, Koura2005Prog.Theor.Phys., Pearson1996Phys.Lett.B} and microscopic models based on non-relativistic~\cite{Goriely2009Phys.Rev.Lett., Goriely2009Phys.Rev.Lett.a, Erler2012Nature} and relativistic density functional~\cite{Geng2005Prog.Theor.Phys., Xia2018Atom.DataNucl.DataTables, Afanasjev2013Phys.Lett.B, Yang2021Phys.Rev.C, Zhang2022Atom.DataNucl.DataTables, Pan2022Phys.Rev.C, Guo2024ADNDT}.
Recently, machine-learning approaches have been widely employed in nuclear mass predictions,  e.g., the kernel ridge regression~\cite{Wu2020Phys.Rev.C051301, Wu2021Phys.Lett.B, Wu2022Phys.Lett.B137394, Guo2022Symmetry, Du2023Chin.Phys.C, Wu2023Front.Phys., Wu2024Phys.Rev.C, Wu2024PRC_AKRR, Guo2024PRC}, the neural network~\cite{Utama2016Phys.Rev.C, Niu2018Phys.Lett.B, Neufcourt2018Phys.Rev.C, Niu2022Phys.Rev.C}, the Gaussian process regression~\cite{Neufcourt2019Phys.Rev.Lett., Shelley2021Universe}, etc.
The machine-learning approaches refine nuclear mass predictions by capturing patterns that may correspond to unperceived physical effects.

Different models include different nuclear effects to different degrees.
Some models may properly consider several of these effects but improperly (less or over) consider several other effects, and some models may be otherwise.
Recently, principal component analysis (PCA) has been employed to extract the principal components (PCs) inherent in various nuclear mass models~\cite{Wu2024SC, Giuliani2024PRR}, which help to understand the major effects that have been captured by present nuclear theoretical models.
It also provides a new strategy to build mass models by reintegrating and reorganizing nuclear effects from different models.

It is also important to investigate residual effects that have not yet been captured by the theoretical nuclear mass models.
In this work, the PCA is employed to extract the PCs of model residuals of various nuclear mass models to investigate unperceived effects of nuclear mass predictions.
The commonalities and differences in model residuals across various nuclear mass models are analyzed using these PCs, aiming to investigate the correlation of effects that are lacking in different nuclear mass models.

\section{Extracting principal components from model residuals}

Principal Component Analysis (PCA) is a statistical technique designed to extract a set of principal components (PCs) that capture the maximum variance within a dataset. 
This method applies a linear transformation that turns the original variables into mutually orthogonal PCs, which are ranked by their eigenvalues (a measure of importance). 
As a result, the first few PCs capture most of the key information, so we can reduce the dimensionality while keeping the essentials.

For the purposes of this paper, we apply PCA to the residuals of several nuclear mass models. 
Here, residuals are the differences between model predictions and the experimental values. Because different models can miss the same pieces of physics, their residuals are often correlated. 
PCA turns these correlated residuals into independent patterns of deviation—what we call principal residual components. 
This makes it easier to identify the physics that these models jointly or solely fail to describe.

To conduct the PCA-based analysis of nuclear mass model residuals, six widely used nuclear mass models, i.e., FRDM2012~\cite{Moeller2016Atom.DataNucl.DataTables}, HFB17~\cite{Goriely2013Phys.Rev.C}, KTUY05~\cite{Koura2005Prog.Theor.Phys.}, D1M~\cite{Goriely2009Phys.Rev.Lett.a}, RMF~\cite{Geng2005Prog.Theor.Phys.}, and LDM~\cite{Weizsaecker1935Z.Physik}, are selected as the source of original model predictions, which are labeled by an index $j$, where $j=1,2,\ldots,N$ and $N=6$.
For each nucleus $i$ and model $j$, we define the mass residual as
$m^{r}_{ij} = m^{\mathrm{exp}}_{i} - m^{\mathrm{th}}_{ij}$,
where $m^{\mathrm{exp}}_{i}$ is the experimental mass taken from AME2020~\cite{Wang2021Chin.Phys.C}, and
$m^{\mathrm{th}}_{ij}$ is the corresponding theoretical prediction from model $j$.  
The overlap of the six selected mass models covers a total of 6254 nuclei. 
Meanwhile, the overlap of these six mass models with the experimental data from AME2020~\cite{Wang2021Chin.Phys.C} includes 2421 nuclei, forming a 2421-dimensional vector in a Hilbert space, where each dimension represents the residual value for a specific nucleus in the nuclear chart.
All model residuals are therefore evaluated on the same $n$ nuclei, so each residual vector has the same dimension $n=2421$.
Consequently, for each model $j$, the residuals form an $n$-dimensional vector
$\bm m^{r}_{j}\in\mathbb{R}^{n}$, whose $i$-th component is $m^{r}_{ij}$. These residual vectors are stacked to form the matrix:
\begin{equation}
X
= \big[\, \bm {m}^{r}_{1}\ \bm {m}^{r}_{2}\ \cdots\ \bm {m}^{r}_{N} \,\big]
=
\begin{bmatrix}
m^{r}_{11} & m^{r}_{12} & \cdots & m^{r}_{1N} \\
m^{r}_{21} & m^{r}_{22} & \cdots & m^{r}_{2N} \\
\vdots     & \vdots     & \ddots & \vdots     \\
m^{r}_{n1} & m^{r}_{n2} & \cdots & m^{r}_{nN}
\end{bmatrix}
\in \mathbb{R}^{\,n\times N},
\end{equation}
where rows index nuclei and columns index models.
To better preserve the physical variances of the nuclear mass residuals, we perform column centering:
for each model $j$ (i.e., column $j$), we first compute its mean residual over the $n$ nuclei, $\bar m^{\,r}_j=\frac{1}{n}\sum_{i=1}^n m^{r}_{ij}$, and then subtract this mean from every entry in that column, i.e., replace $m^{r}_{ij}$ by $m^{r,\mathrm{cen}}_{ij}=m^{r}_{ij}-\bar m^{\,r}_j$.
For notational simplicity and ease of understanding, we continue to denote the centered matrix by $X$ and centered residual vector by $\bm m^{r}_{j}$. 
On this basis, we construct the covariance matrix:
\begin{equation}
C = \frac{1}{n-1} X X^\top \in \mathbb{R}^{\,n\times n}.
\end{equation}
This covariance matrix $C$ captures the co-variation of model residuals across nuclei, and its eigenvectors $\bm v_k\in\mathbb{R}^{n}$ form a set of orthogonal basis in the $n$-dimensional nuclear space. 
We achieve the diagonalization of the covariance matrix by solving its eigenvalue equation:
\begin{equation}
  C\,\bm v_k = \lambda_k\,\bm v_k,
\end{equation}
where $\lambda_{k}$ are the eigenvalues. 
Since the rank of  covariance matrix $C$, $\mathrm{rank}(C)\le \min(n,N)=6$, at most six eigenvalues are nonzero. 
Their corresponding eigenvectors $\bm v_{k}$ form the extracted principal components and are visualized on the nuclear chart in Fig.~\ref{fig2}; we will discuss them below. 
The principal components related to the six nuclear mass model residuals are labeled as PC1, PC2, ..., and PC6.

After obtaining the principal components, for each model $j$ we project its centered residual vector $\bm m^{r}_{j}$ onto every principal direction $\bm v_k$.
This projection quantifies the contribution of that PC pattern to a model’s residuals. 
In physical terms, the projection measures how much of each independent error mode the model inherits, revealing which shared missing physics dominates its deviations.
Finally, the overlap (cosine similarity) between each residual vector $\bm m^{r}_{j}$ and principal component $\bm v_{k}$ can be easily calculated by
\begin{equation}
a_{jk}=\frac{\bm m^r_j \cdot \bm v_k}{\|\bm m^r_j\|\,\|\bm v_k\|}\in[-1,1],
\end{equation}
which measures how strongly the residuals of model $j$ follow the $k$-th PC of mass model residuals. 
Values near $+1$ (or $-1$) indicate strong in-phase (or out-of-phase) correspondence with the PC pattern, whereas values near $0$ indicate a weak relation.

\section{Results and discussions}

In the following, we explore the underlying structure and physical implications encoded in the residual correlations among different nuclear mass models. 
Rather than focusing on individual numerical deviations, our analysis aims to extract the dominant patterns that characterize systematic trends shared across models as well as the distinct deficiencies specific to each approach. 
By decomposing the residual patterns into an orthogonal basis of principal components, we identify the key directions in the model space along which the largest variations occur.
This provides a compact representation of the residual landscape and reveals the physical origins of model correlations in terms of shell effects, deformation behavior, odd-even effects, and even unknown dependencies. 
The discussion below connects these statistical features to interpretable physical mechanisms and tries to assess how such insights can guide future refinement of global mass models.

\begin{table*}[!ht]
\centering
\caption{Corresponding eigenvalues of the six principal components extracted from the six nuclear mass model residuals. 
The overlaps of the six corresponding principal components (PCs) with the six nuclear mass model residuals.
The eigenvalue of PC1 is scaled to 1, with all other principal components undergoing corresponding scaling.}
\setlength{\tabcolsep}{15pt} 
\renewcommand{\arraystretch}{1.2} 
\begin{tabular}{l|cccccc}
\hline
 Models & PC1 & PC2 & PC3 & PC4 & PC5 & PC6 \\
\hline
Eigenvalues & $1$ & $5.6\times10^{-1}$ & $4.7\times10^{-1}$ & $3.1\times10^{-1}$ & $3.0\times10^{-1}$ & $2.7\times10^{-1}$\\
FRDM2012 & 0.7223 & 0.0681 & -0.0549 & 0.6194 & -0.2923 & -0.0405 \\
HFB17 & 0.5677 & -0.3346 & -0.5209 & -0.3985 & -0.3621 & -0.0669 \\
KTUY05 & 0.7188 & 0.3112 & -0.1164 & -0.1045 & 0.2905 & 0.5269 \\
D1M & 0.6794 & -0.3881 & 0.1138 & -0.0041 & 0.4880 & -0.3697 \\
RMF & 0.4325 & -0.0468 & 0.8212 & -0.2382 & -0.2769 & 0.0545 \\
LDM & 0.2279 & 0.8819 & -0.0587 & -0.1766 & -0.0173 & -0.3680 \\
\hline
\end{tabular}
\label{tab1}
\end{table*}

The eigenvalues corresponding to the six PCs from nuclear mass model residuals are presented in Table~\ref{tab1}, together with overlaps between the six nuclear mass model residuals and the six PCs.
The eigenvalue represents the importance of the corresponding principal component.
As can be seen in Table 1 of Ref.~\cite{Wu2024SC}, the eigenvalue related to the first principal component is overwhelmingly dominant.
As mentioned in Ref.~\cite{Wu2024SC}, this indicates the large similarity among different nuclear mass models.
This similarity can be seen from the overlaps of the six nuclear mass models with the first principal component in Table 1 of Ref.~\cite{Wu2024SC}, which are similar and near 0.999.
The first principal component thus represents the common feature included in different nuclear mass models.
Inspection of the other principal components reveals differences with the first principal component.
As seen in Table 1 of Ref.~\cite{Wu2024SC}, their eigenvalues are much smaller than that of the first principal component, and their overlaps with different mass models are relatively small and no longer similar to each other.
These principal components represent the features that contribute to the differences among nuclear mass models.

\begin{figure}[!htbp]
  \centering
  \includegraphics[width=0.95\linewidth]{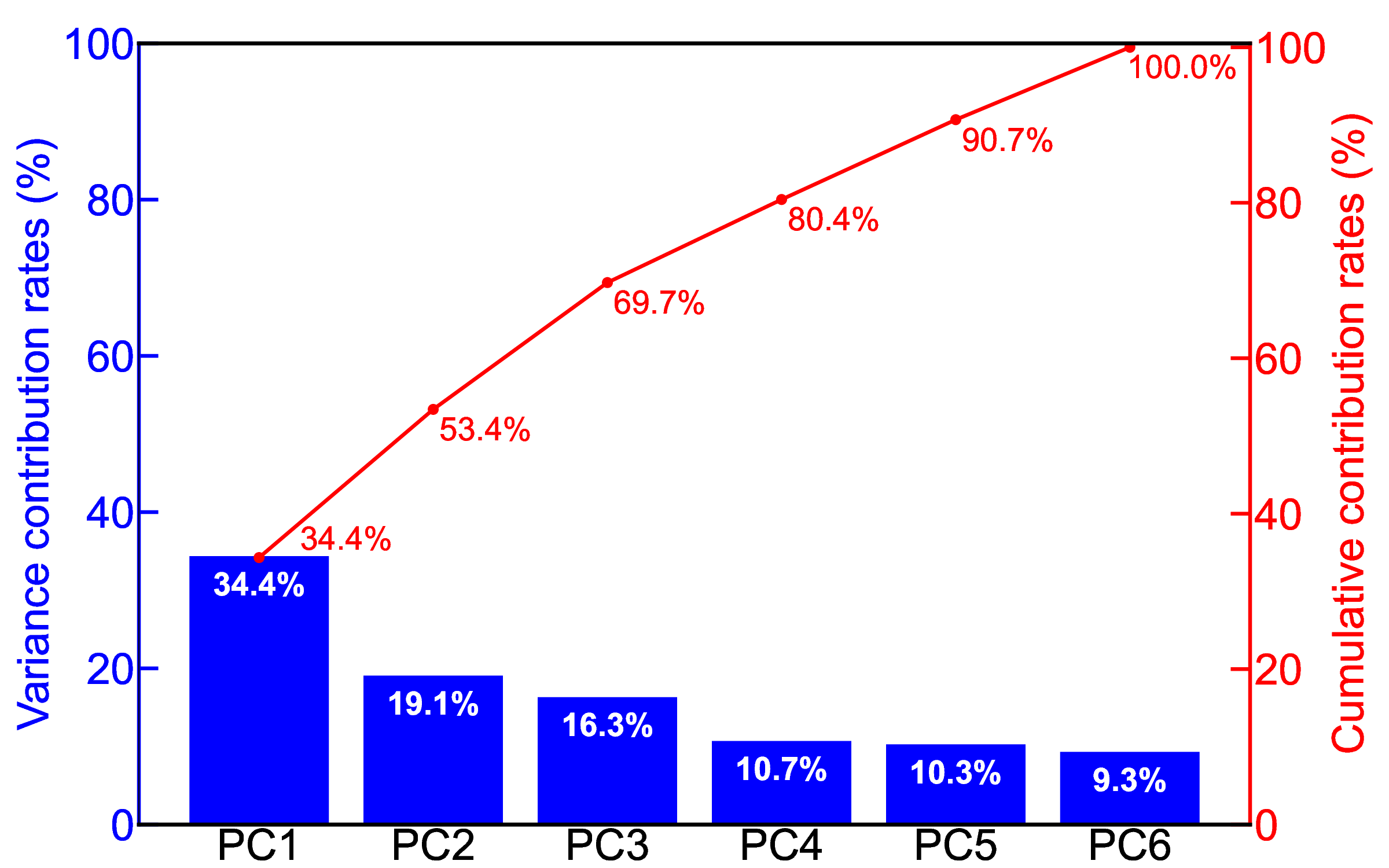}
  \caption{Variance contribution rates (blue bars) and cumulative contribution rates (red line) of principal components for nuclear mass model residuals.
  The variance contribution rate is an important concept in the principal components analysis, which represents the contribution rate of each principal component in the representation of the models.
  }
  \label{fig1}
\end{figure}

Things are different for the PCs from different nuclear mass model residuals.
As can be seen in Table~\ref{tab1}, the dominance of the eigenvalue related to PC1 is far less pronounced, although it is still the largest by definition.
One can also see that the overlaps between PC1 and nuclear mass model residuals are comparable to those of other PCs.
Especially, for the residuals of LDM and RMF models, the dominant principal components are PC2 and PC3 respectively instead of PC1.
This indicates a low similarity among the residuals of different nuclear mass models; that is, the residuals of various models are quite distinct from each other.
It is not good news that no single PC dominates the features contained in the various nuclear mass model residuals; otherwise, we could pinpoint what features the nuclear mass models are missing in general via analysing this principal component, and then use this insight to enhance the description of nuclear masses.
The contribution rates of these PCs from various nuclear mass model residuals are presented in Fig.~\ref{fig1}.
One can see that the first principal component (PC1) contributes 34.4\% of the residual variance across models.
However, the contribution rates of the other PCs, ranging between 10\% and 20\%, are comparable to that of PC1.
This means that all these PCs are important residual features of nuclear mass models, and no single principal component can dominate the residual patterns.

\begin{figure*}[!htbp]
  \centering
  \includegraphics[width=0.95\linewidth]{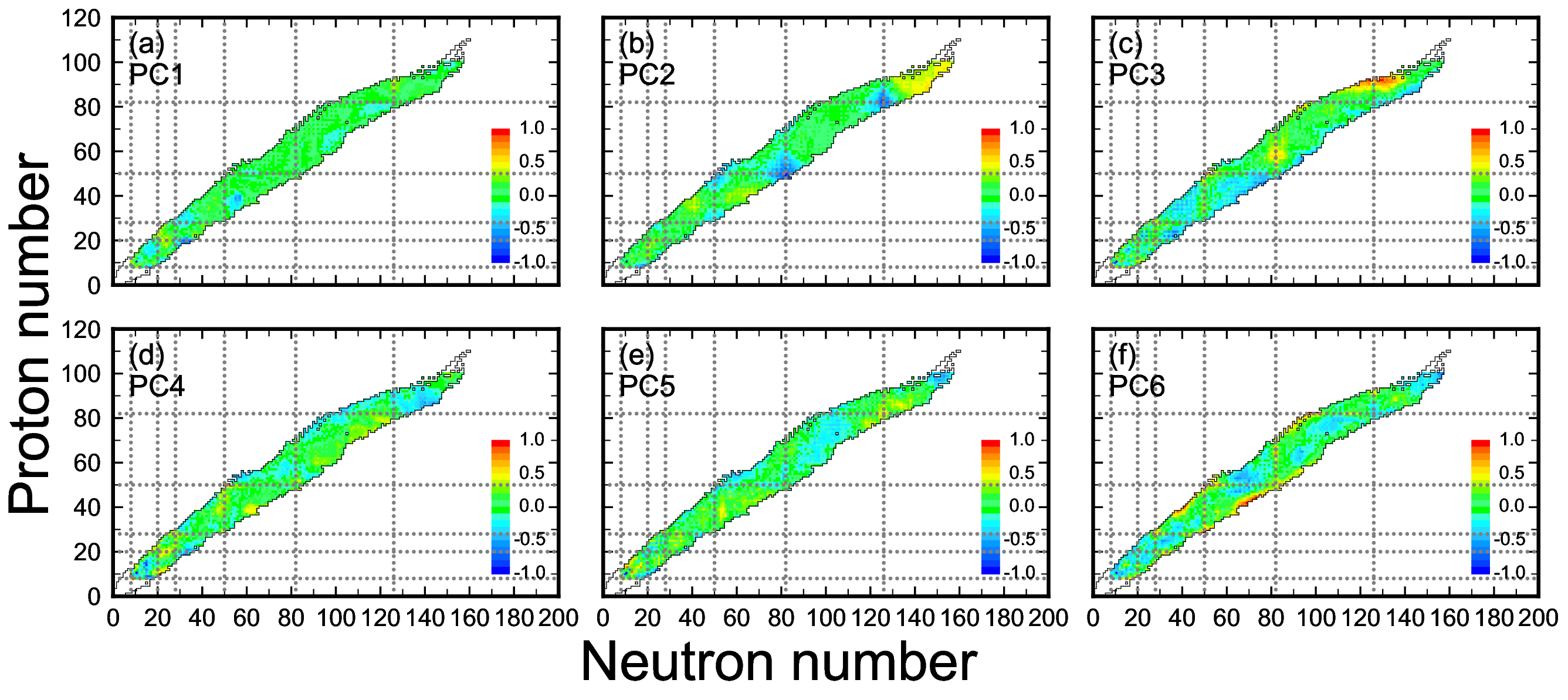} 
  \caption{Principal components, i.e., PC1 (a), PC2 (b), ..., and PC6 (f), of nuclear mass models with the values scaled to the range between -1 and 1. The boundary of nuclei with known masses in AME2020 is shown by the black contour lines. Dotted lines indicate the magic numbers.}
  \label{fig2}
\end{figure*}

Principal components (PCs) from six nuclear mass model residuals are presented in Fig.~\ref{fig2}.
They represent the principal patterns encoded in the residuals of these six nuclear mass models.
The hope is that one can obtain some information from these PCs, so that one can find a proper way to reveal the remaining information not captured by nuclear mass models.
Specifically, if certain principal components show obvious patterns, it indicates that there are systematic deviations in nuclear mass models in those aspects, which can guide the improvement of the nuclear mass models.

Since no single principal component dominates, one cannot find general features that would be important to enhance various nuclear mass models.
One can only suggest for each nuclear mass model some features from specific one or two PCs. 
The contributions of the six PCs to each nuclear mass model residual are presented in Table~\ref{tab1}.
Since the PCs form an orthonormal basis, the overlap is the direct projection of a model’s residual pattern onto a given PC, so a larger overlap magnitude indicates that this PC captures a larger fraction of the model’s systematic residual structure.

Table~\ref{tab1} indicates that PC1 is the most important feature for four nuclear mass model residuals, i.e., FRDM2012, HFB17, KTUY05, and D1M.
This suggests that for well-designed nuclear mass models, the principal residual patterns may be similar, which could be helpful for further refining these four models.
In Fig.~\ref{fig2}(a), PC1 exhibits fine-grained structures concentrated in the light-nuclei region, reflecting mild odd--even and shell-related features there.
According to Table~\ref{tab1}, PC2 is extremely important for the residuals of the LDM model and is also important for the HFB17, KTUY05, and D1M models.
The prominent feature of PC2 is the deformation properties related to shell effects, as illustrated in Fig.~\ref{fig2}(b) with the help of the magic lines.
This implies that the LDM model lacks shell effects, which is already well known.
It also indicates that residual shell effects have not been fully incorporated into the HFB17, KTUY05, and D1M models.
Table~\ref{tab1} further shows that PC3 is important for the RMF and HFB17 models.
As illustrated in Fig.~\ref{fig2}(c), PC3 includes features related to shell structures, odd-even behaviors, and superheavy nuclei.
Since HFB17 and RMF are the adopted microscopic nuclear mass models, these features may be important for further refining microscopic models.
Moreover, PC4 is important for FRDM2012, PC5 for D1M, and PC6 for KTUY05 (see Table~\ref{tab1}).
These PCs exhibit subtle patterns with wide and fragmented distributions across the nuclear chart, indicating fine-scale structures that likely reflect a superposition of multiple effects.
Such components are difficult to attribute unambiguously to a single known physical mechanism, and they may include subtle patterns of known effects as well as possible unknown effects.

These PCs of nuclear mass residuals can be employed to improve nuclear mass predictions of each model respectively.
This correction is achieved by projecting the corresponding most important principal component onto the original residual pattern for a specific nuclear mass model, and then adding this part of principal component to the model.
The optimized nuclear mass model predictions can be written as
\begin{equation}
    \bm M^{\mathrm{opt}}_j = \bm M^{th}_j + \Delta\bm M_{jk},
\end{equation}
where $\bm M^{th}_j$ denotes the theoretical mass predictions from model $j$, and $\Delta \bm M_{jk}$ is the correction related to the most important principal component, 
\begin{equation}
    \Delta \bm M_{jk}=\left( \frac{\bm m^r_j \cdot \bm{v}_k}{\|\bm{v}_k\|^{2}} \right) \bm{v}_k.
\end{equation}

\begin{figure}[!htbp]
  \centering
  \includegraphics[width=0.95\linewidth]{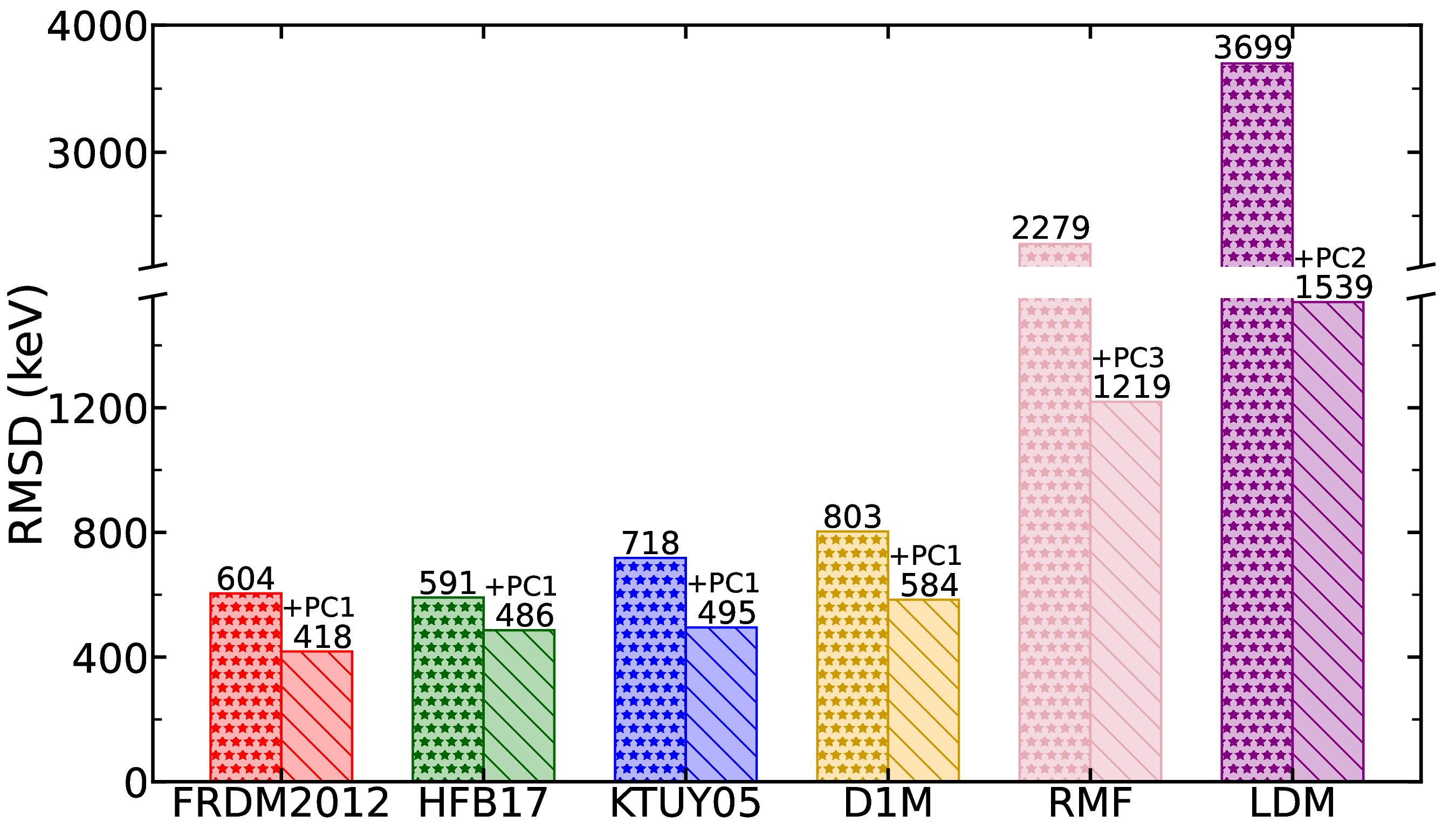} 
    \caption{
    Root-mean-square deviations (RMSDs) of six nuclear mass models with respect to the AME2020 experimental masses (left bar in each model group), compared with the ones optimized by adding the most important principal component (right bar).
    The principal component used for each optimization is indicated above the corresponding right bar.
    Numeric RMSD values are printed on top of the bars.
    A broken $y$-axis is employed to display simultaneously the comparatively small RMSDs of the first four models and the larger RMSDs of RMF and LDM.
    }
  \label{fig3}
\end{figure}

The root-mean-square deviation (RMSD) relative to the AME2020 experimental masses is employed as a global metric to evaluate the accuracy of nuclear mass models.
The RMSDs of six nuclear mass models, and their corresponding optimized versions are presented in Fig.~\ref{fig3}.
These optimized versions are obtained by incorporating the related most important PC into the residual pattern of each original model respectively.
The most important PC adopted for each model is selected according to the largest overlap as shown in Table~\ref{tab1}.
As shown in Fig.~\ref{fig3}, correcting each model with only the most important PC can reduce the RMSD significantly.
This indicates that a substantial portion of the residual in each mass model is not random noise but rather follows a systematic pattern, which can be effectively captured by a single PC.
PC1 is the most important residual pattern for four nuclear mass model residuals, i.e., FRDM2012, HFB17, KTUY05, and D1M, which helps to refine these four nuclear mass models significantly.
This indicates that these four well-calibrated global models share a similar dominant deficiency.
This dominant pattern is mainly related to mild odd-even staggering and shell-associated features in the light-nuclei region as shown in Fig.~\ref{fig2} (a).
The optimal PC differs for the remaining two models, consistent with their different characteristics of residual patterns.
As discussed, the prominent feature of PC2 is the deformation properties related to the shell effects, which is precisely what the LDM model lacks the most.
It therefore works very well to improve the prediction accuracy of the LDM model by adding PC2.
For the RMF model, PC3 provides the largest reduction, lowering the RMSD from 2279 to 1219~keV, implying that the dominant RMF deficiency is closely related to the residual features encoded in PC3.
As discussed in Fig.~\ref{fig2} (c), PC3 includes features related to shell structures, odd-even behaviors, and superheavy nuclei.
Finally, the remaining RMSD after the single-PC optimization implies that additional residual structures remain and merit further investigation.

\section{Summary}

This study employs Principal Component Analysis (PCA) to extract and analyze the principal components of model residuals from six widely used nuclear mass models, aiming to explore unperceived effects missing in current nuclear mass predictions. 
Key findings reveal distinct characteristics compared to the principal components of nuclear mass models themselves.
No single principal component of the nuclear mass model residuals takes an overwhelmingly dominant role, which means that residuals of different models are largely uncorrelated, indicating current nuclear mass models fail to account for some underlying physical mechanisms in diverse ways.
 
Further analysis links specific 
PCs to model-specific missing effects: PC1 is critical for refining FRDM2012, HFB17, KTUY05, and D1M, with features concentrated in light nuclei; PC2 (related to shell effects) is vital for LDM and relevant for HFB17, KTUY05, and D1M; 
PC3 (involving some fine features related to shell structures, odd-even behaviors, and superheavy nuclei) matters for microscopic models RMF and HFB17. 
PC4, PC5, and PC6, while important for FRDM2012, D1M, and KTUY05 respectively, exhibit subtle patterns with wide and fragmented distributions across the nuclear chart, indicating they include fine-scale structures that likely reflect a superposition of multiple effects.
These components are difficult to attribute unambiguously to a single known physical mechanism.
They should include some subtle patterns of known effects and even some unknown effects.

In conclusion, the lack of a universally dominant residual pattern suggests that there is no single missing ingredient that can uniformly improve all existing nuclear mass models. 
Instead, model improvements should be guided by the specific principal components most relevant to each model residual. 
This work provides a data-driven strategy to identify model deficiencies and guide the refinement of nuclear mass predictions.

%===============Acknowledgments ==============

\section*{Acknowledgments}

This work was partly supported by the National Natural Science Foundation of China (Grant No. 12405134), the China Postdoctoral Science Foundation (Grant No. 2021M700256), and the start-up grant XRC-23103 of Fuzhou University.

\bibliographystyle{elsarticle-num}
\bibliography{paper}

%\clearpage

%%-----------------------------------------------------
%\section*{Appendix}
%%-----------------------------------------------------

\end{document}